# Astro2020 Science White Paper

# Stellar Characterization Necessary to Define Holistic Planetary Habitability

**Thematic Areas:** ☒ Planetary Systems ☒ Star and Planet Formation
☐ Formation and Evolution of Compact Objects ☐ Cosmology and Fundamental Physics
☒ Stars and Stellar Evolution ☐ Resolved Stellar Populations and their Environments
☐ Galaxy Evolution ☐ Multi-Messenger Astronomy and Astrophysics

**Principal Author:**
Name: Natalie Hinkel
Institution: Southwest Research Institute
Email: natalie.hinkel@gmail.com

**Co-authors:** Irina Kitiashvili [BAERI at NASA ARC], Patrick Young [Arizona State University], Allison Youngblood [NASA GSFC]

**Co-signers:** Vladimir Airapetian [NASA GSFC/SEEC and American University], Vardan Adibekyan [IA, UP, Porto, Portugal], Benjamin V. Rackham [University of Arizona]

**Abstract**
It is a truism within the exoplanet field that "*to know the planet, you must know the star*." This pertains to the physical properties of the star (i.e. mass, radius, luminosity, age, multiplicity), the activity and magnetic fields, as well as the stellar elemental abundances which can be used as a proxy for planetary composition. In this white paper, we discuss important stellar characteristics that require attention in upcoming ground- and space-based missions, such that their processes can be understood and either detangled from that of the planet, correlated with the presence of a planet, or utilized in lieu of direct planetary observations.

**Introduction**
Stars and planets are inextricably linked, both physically and chemically, in such a way that the star directly influences whether or not the planet is habitable (see Fig 1 for illustrative flow chart). Stars that are too young will have unstable planets that are still forming; older main sequence and post-MS stars will have had their luminosities increase and their radii expand, increasing irradiances and potentially engulfing close-in planets. Accurate stellar masses and radii are required for measuring planetary masses and radii. Stellar luminosity, and the distance between the planet and its host, defines the habitable zone, or the orbital region where we would expect to find liquid water and temperate planetary surface temperatures for planets with an adequate atmosphere. The chromospheric activity of a star, ultimately tied to stellar rotation through magnetic dynamo processes and usually a strong indicator of stellar age, indicates the strength of stellar magnetic fields and impacts the frequency and power of stellar flares and coronal mass ejections -- phenomena which strongly impact planetary atmospheres. In addition to these more physical attributes, stars and planets are formed at the same time, out of the same molecular cloud. This means that their basic, elemental compositions are related to one another -- a relationship that is vital until such time as we are able to directly measure the composition of an exoplanet. For all of these reasons, it is necessary to precisely and accurately characterize the physical and chemical properties of a planet's host star such that its relationship with the planet, and overall planetary habitability, can be assessed.

**1. Improving Stellar Models**
Stars are the backdrop against which we are able to detect and characterize an exoplanet. For proper characterization of planets it is critical to understand physical properties and dynamics of the host stars. Inferences of stellar mass (for single stars without dynamical mass estimates) and age often come from comparison of stellar parameters (e.g. temperature, luminosity, surface gravity, chemical abundances, asteroseismic data, etc.) to theoretical evolutionary tracks and isochrones. Until recently, only 1D stellar models based on mixing-length theory have been used. At the present time, fast growing computational capabilities allow us to generate 3D time dependent simulations of stellar convection that are capable of reproducing the surface turbulent dynamics and radiation properties with a high degree of realism. These type of models more realistically take into account the EOS, stellar interior structure, composition, and effects of radiation, magnetic fields, and turbulence.

Granular flickering, convective blueshift, gravitational redshift variations, and magnetic fields cause fluctuations of spectral line centroids and introduce a correlated signal into precision radial velocity (PRV) measurements used to discover and measure masses of exoplanets. However, as it is known from solar observations, surface convection phenomena involve many scales which cannot be considered with 1D models. Based on first physical principles, 3D radiative magnetohydrodynamic (MHD) models have been actively used to accurately model solar surface convection and study many phenomena on the Sun (e.g., Stein et al., 2011; Chen et al., 2017). Similar approaches can be used to obtain 3D dynamical stellar models, the initial conditions for which are computed using a stellar evolution code to fit the observed stellar properties (mass, radius and composition).

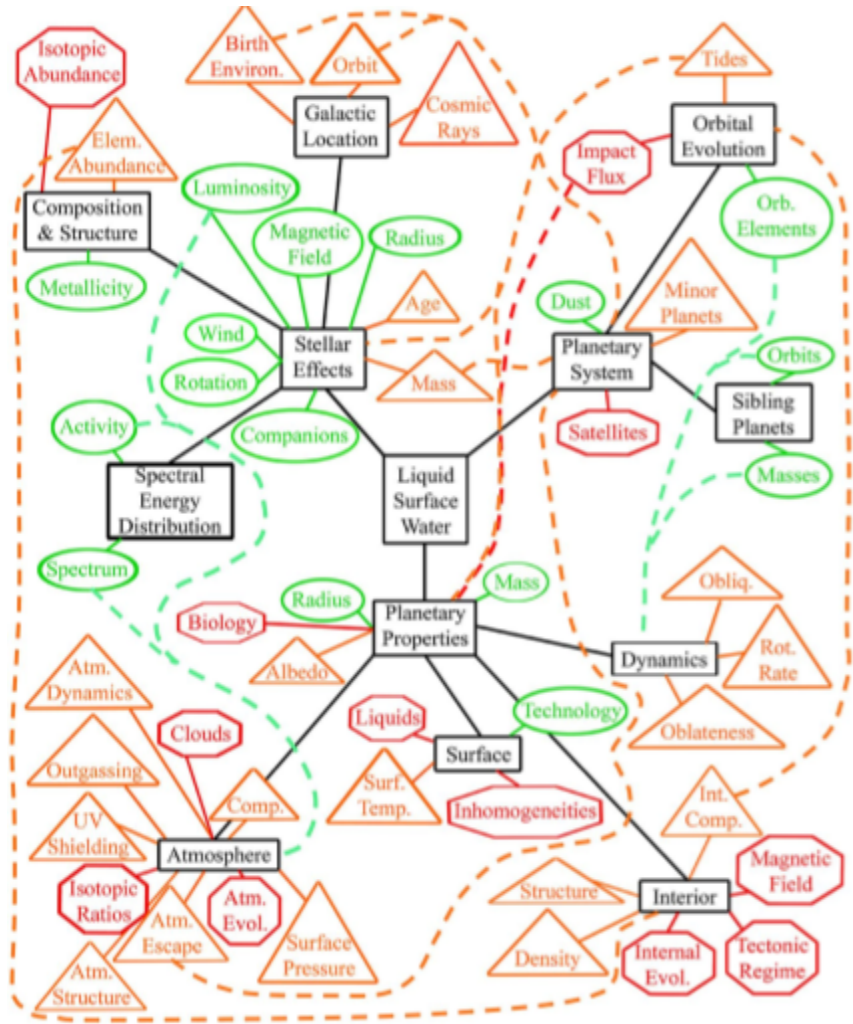

Fig 1. Flow chart illustrating the how stellar and planetary properties are connected to one another. Green ovals are characteristics that are directly observable, orange triangles can be inferred from models, and red hexagons are theoretical. Courtesy of Rory Barnes and Vikki Meadows at VPL.

Obtained 3D models can be used to characterize more precisely stellar properties such as abundances (Section 2), determine characteristic scales of the stellar granulation, investigate effects of magnetic fields and rotation on internal structure and dynamics of the host stars, and generate synthetic spectral and spectropolarimertic data series. Currently available 3D stellar simulations (e.g., Kitiashvili et al., 2016) have demonstrated variations of the granulation scales for different type of main-sequence stars, as well as significant deviations of stellar characteristic (e.g., thermodynamical properties, radius) from the 1D mixing-length models, which must be taken in to account for characterization of planet-hosting stars. Fortunately, evolutionary codes exist with sophisticated treatments of convective hydrodynamics that correct many of the shortcomings of mixing length models (e.g. Truitt et al. 2015). These can be used as background conditions for 3D atmospheres, and making the shift to such models would benefit evolutionary characterization. Future development of 3D radiative hydrodynamic and MHD models of the stellar turbulent dynamics is required along with development of techniques for characterization of stellar parameters, and is necessary for cross-validation of both models and observational data analysis methods.

**2. Stellar Abundances**
After mass, chemical composition plays the largest role in determining how stars evolve. Our knowledge of stellar abundances must encompass both bulk abundance of elements heavier than H and He (metallicity) and specific abundance ratios. For the purposes of astrobiology, this translates into how much and how quickly the insolation and incident stellar spectrum at a given orbital distance changes. These properties are primary determinants of the location of the classical habitable zone. Predicting a planet's climate history and duration in the habitable zone are important considerations in assessing the likelihood that it could produce and sustain detectable biosignatures under the assumption that long-term environmental stability is important to sustaining a productive ecosystem (e.g. Dong et al. 2018). As an example, the range of O/Fe abundances observed within 100pc of the sun can result in a variation of main sequence lifetimes of ~3Gy for a solar mass star at solar [Fe/H]. Using a radiative-convective prescription for the location of the inner and outer edges of the habitable zone, the habitable lifetime of a planet in a 1AU orbit around the host star ranges from ~3.5 - 8 Gy (Truitt et al. 2015).

Beyond controlling the evolution of the star itself, stellar abundances provide the most accessible and often the only information on the planetary composition in the system. Limited information on atmospheric composition can be gained from transit spectroscopy -- with only ~20 exoplanetary atmospheres directly measured. Indeed, atmospheric composition has yet to determined for an Earth-sized planet in the habitable zone, due to detection and geometric constraints. However, since stars and planets are formed at the same time from the same bulk materials within the molecular cloud, the composition of the star can be used as a proxy for the interior makeup of the planet. For example, Thiabaud et al. (2015) analyzed whether the C, O, Mg, Si, and Fe were the same within the planet as a star, using a planet formation and composition model, and found that these important rock-forming elements were the same in both the star and planet. To this end, a variety of models have utilized stellar abundances to constrain the interiors of terrestrial exoplanets, i.e. ExoPlex -- which determines the mineralogy and density of planetary interiors (Unterborn et al. 2016, 2018a/b, Hinkel & Unterborn 2018) and a Bayesian-based probabilistic inverse analysis (Dorn et al. 2017). The overall, holistic habitability of an exoplanet is dependent on its surface conditions, internal structure, mineralogy, and atmosphere (Foley & Driscoll 2016), where we must use host star measurements to understand the majority of these properties.

**3. Activity and Magnetic Fields**
Stellar magnetic activity manifests in starspots, X-ray and UV emission, flares, and coronal mass ejections, which can each affect exoplanet detection and characterization efforts and/or physically alter the exoplanets themselves. Magnetic activity is due to rotation and convection, and appears to occur throughout the late-A, F, G, K, and M spectral classes, even for late-type M dwarfs with fully-convective interiors (e.g., no tachocline).

For exoplanet characterization efforts with transmission spectroscopy, characterizing the spot coverage fraction and any time variability is critical due to the transit light source effect (Rackham et al. 2018, 2019). Photospheric heterogeneity (e.g., bright or dark spots) can mask or mimic planetary signals, particularly for later spectral types (K and M). Simultaneously obtaining visual transmission spectroscopy with longer-wavelength spectroscopy can help mitigate this effect (see white paper ``Constraining Stellar Photospheres as an Essential Step for Transmission Spectroscopy of Small Exoplanets'' by B. Rackham for more details).

For biosignature searches or photochemical modeling of any kind of planet, characterization of the absolute flux and spectral properties of the host star's UV emission must be pursued (Domagal-Goldman et al. 2014; Harman et al. 2015). Direct observations are recommended, as self-consistent stellar models that treat the chromosphere, transition region, and corona are not available, and current models are not readily applicable to multiple stars (Fontenla et al. 2016 for GJ 832). Empirical scaling relations have been developed to estimate UV emission from easier-to-observe emission, including Lyman alpha (Shkolnik et al. 2014; Youngblood et al. 2017), but it is not quantified how uncertainties in these estimates impact exoplanet photochemical models.

Stellar variability due to flares is most pronounced at X-ray and UV wavelengths (e.g., factors of a few to several orders of magnitude) where emission from magnetically-heated regions dominates. Ultra-precise photometry from *Kepler* and *TESS* in the optical can catch flares, but the absence of flares in the optical does not mean that the star is quiet in the X-ray and UV, whose impact on exoplanet atmospheres is pronounced. Loyd et al. (2018a,b) indicate that emission from flares may actually dominate the UV emission of M dwarf stars, so the impact of UV flares on exoplanet photochemistry could be tremendous. Characterizing or simply detecting coronal mass ejections or energetic particles (often associated with large solar flares) without *in situ* detectors or coronagraphs (with inner working angles <1 mas) is challenging. The path forward may lie in the EUV (100-912 Å) where coronal emission lines dim by detectable amounts after part of the quiet corona is evacuated by a coronal mass ejection (Mason et al. 2014; see white paper ``EUV observations of cool dwarf stars'' led by A. Youngblood).

**4. Stellar Ages**
When prioritizing targets for follow-up biosignature observations, dwell time in the habitable zone may be an important consideration, as may the heat flux and potential for geological activity of an exoplanet. These considerations make the stellar age an essential piece of information. The ages of low mass field stars are in most cases difficult to determine to precision of better than a gigayear. Techniques for more precisely determining ages (gyrochronology, age-activity relations, lithium depletion, surface gravity features, metallicity, and kinematics) are far more effective for stars younger than 1 Gy (e.g. Mamajek & Hillenbrand 2008; Almeida-Fernandes & Rocha-Pinto 2018). These indicators are generally weak or poorly correlated at old ages, so more reliance must be placed on ages from evolutionary tracks. The slow evolution of low mass stars requires precision knowledge of mass and composition of individual stars for initial conditions of models along with high precision measurements of luminosity and effective temperature and/or radius for fitting to evolutionary tracks. Observationally the two most significant limitations on the precision of ages for older field stars more than ~0.8 solar masses are distance and bolometric magnitude, both contributing to uncertainties in luminosity. The advent of Gaia has greatly reduced the uncertainty in distance for the vast majority of potential target stars. Intrinsic colors of stars on the pre-MS and along the MS vary for the same spectral type due to reddening, stellar activity, and the frequent reliance on synthetic IR colors (e.g. Casagrande et al. 2008). Precision effective temperatures or radii are increasingly important for lower mass stars. High resolution spectroscopy is the most reliable method of obtaining $T_{eff}$, but still suffers from systematic uncertainties (e.g. Hinkel et al. 2016). A program of high resolution optical and IR spectroscopy, especially combined with improved techniques for interpreting spectra of M stars, and asteroseismology will ameliorate uncertainties in both bolometric luminosity and radius/$T_{eff}$.

**Conclusion**
There are stellar properties that contribute directly to exoplanet detection (e.g. stellar mass and radius) and those that impact confirmation (e.g. starspots). There are characteristics of the star that directly influence a planet's surface, such as activity, coronal mass ejections, and X-rays. And there are attributes of the star that cannot currently be measured within the planet, such as elemental abundances. The composition of the planet is absolutely fundamental to whether the planet can have active geochemical cycles, feeding new elements to different regions of the planet, including the atmosphere. Whether direct or indirect, the stellar host deeply influences the orbiting planet. Therefore, it is important that, as a community, we focus on the holistic definition of planet habitability. It is not simply enough to have a planet lie within the liquid water Habitable Zone. We must also ask: Is there water on the surface of the planet? How long has the water been maintained? Could life exist or is the star too active for life to survive? Are the elements necessary for life present or in the planet? Upcoming missions -- such as JWST and WFIRST -- will be focusing on finding and characterizing small, Earth-sized planets in order to find an Earth-analogue. It is vital that we prepare in advance, by observing and measuring the properties for the target stars that will directly affect the habitability of the planet, both physically and chemically. Therefore, we recommend:
- Support for access to high resolution optical and near-IR spectrographs on large telescopes to measure precise stellar parameters and chemical abundances, and ability to measure stellar spectra throughout the UV and X-rays.
- Support for data analysis for determination of precise stellar properties for exoplanet host stars and exoplanet survey targets.
- Support for theoretical astrophysics research on improving models of stellar interiors, atmospheres, and evolution, which incorporate the latest advances in computing, EOSs, opacities, nuclear networks, MHD.